\documentclass[]{pasj02} 
\usepackage[switch,mathlines]{lineno} 

\jyear{2024}
\Received{}
\Accepted{}

\begin{document} 

\title{ Orbital Contraction of Post-Common-Envelope Binaries with a Circumbinary Disk }

\author{
 Shigeyuki  \textsc{Karino}\altaffilmark{1}\altemailmark\orcid{0000-0002-1346-2242} \email{karino@ip.kyusan-u.ac.jp} 
 and
 Kenji \textsc{Nakamura}\altaffilmark{1}\orcid{0000-0001-8047-6743}
}
\altaffiltext{1}{Faculty of Science and Engineering, Kyushu Sangyo University, 2-3-1 Matsukadai, Higashi, Fukuoka, Japan}



\maketitle

\begin{abstract}

Tight and compact binary systems, such as double neutron star binaries, are believed to undergo a common envelope evolution phase, resulting in strongly bound orbits. 
During this phase, the outer layers of the primary star are expelled, resulting in orbital shrinkage. 
However, a part of the expelled material may remain as a circumbinary disk, which can further influence subsequent orbital evolution. 
In this study, we investigated orbital evolution in the presence of a circumbinary disk within a simplified framework by assuming that orbital contraction and disk dissipation occur over the viscous timescale. 
The results showed that the orbit of the binary system after the common envelope evolution phase was further contracted by up to $\sim 17 \%$ due to the presence of the circumbinary disk, irrespective of the disk's mass and structure. 
This additional orbital contraction following the common envelope evolution phase may have significant implications for the formation rate of double neutron star binaries that merge within a cosmic timescale.

\end{abstract}


\section{Introduction}

The presence of a circumbinary disk (CBD) surrounding a binary system enables the transfer of the orbital angular momentum of the binary to the disk through tidal interactions, thereby contracting the orbit of the binary system.
Such CBD-induced orbital contraction has been thought to accelerate the merger process of black holes.
The presence of a CBD around a black hole binary at a galactic center is expected to accelerate the merging of the black holes and may contribute to the growth of supermassive black holes \citep{AN02, BBR80, IPP99}.

CBDs are also likely to be formed around binary systems after the occurrence of non-conservative and drastic mass transport processes such as common envelope (CE) evolution \citep{S98, DM11, I11, KS11, KAB14}.
In binary systems with short orbital periods, when mass transport becomes unstable, one of the stars (a giant star) may overflow its Roche lobe and form a CE \citep{P76, IL93, I13}.
In this case, the companion star is subjected to large friction during its orbital motion in the CE, resulting in the rapid contraction of its orbit \citep{IL93, S98}.
At this time, the orbital kinetic energy is transferred to the envelope matter and the outer layers of the giant star are blown away.
If the orbit contracts excessively and reaches the central core of the giant star, the binary system merges.
Even if no direct merger occurs during CE evolution, the residual CBD will cause the orbit of the surviving binary system to rapidly shrink due to the interaction between the binary system and the CBD \citep{A91, KS11}.

CE evolution is believed to play an important role in binary evolution.
For example, if a binary system survives through CE evolution, the residual binary system will have a very short orbital period.
At least some cataclysmic variables likely formed in this manner through CE evolution \citep{P76, P84}.
Short-period double white dwarf binary systems formed via CE evolution are attracting attention as the progenitors of type Ia supernovae \citep{IT84, I13}.
Furthermore, binary systems composed of compact objects, which merge within the cosmic age, are believed to undergo orbital contraction during CE evolution \citep{B02, D12, T17}.
Conversely, subdwarf-B stars with unusual surface compositions may result from mass transfers and binary mergers during CE evolution \citep{H02}.
In addition, it has been suggested that the outer layer of the giant expanding due to CE evolution can be observed as luminous red novae \citep{P19, ST03}.

In recent years, three-dimensional numerical hydrodynamic simulations of CE evolution have been conducted, wherein the companion star is treated as a point source spiraling into the outer layers of the giant star \citep{K20, L22, O22, P12, RT12, V24}.
However, the self-consistent three-dimensional modeling of CE evolution covering the entire period from the plunge-in of the companion star to the complete ejection of the envelope is difficult due to the enormous spatial and temporal scales involved in this problem.
In most numerical simulations, the outer layer cannot be fully ejected within the computation duration.
Previous attempts have either been terminated immediately after a rapid spiral-in, leaving a considerable amount of gravitationally bound matter within the system, or have omitted this plunge-in phase entirely and modeled only the subsequent evolution of the post-plunge-in configuration \citep{V24}.
In these studies, numerical simulations have been employed to investigate how the central binary exchanges orbital angular momentum with the CBD \citep{AN02, MSS19}.
In recent years, numerical simulations have also been conducted for studying the orbital evolution of post-CE binary systems, particularly the influence of a CBD on binary evolution \citep{DD21, SWH23}.
These simulations suggest that in some cases, outer layers can form a disk around the binary system and interact with it over long periods \citep{W24}.
In general, such numerical simulations incur high computational costs.
Therefore, the trackable time span is not long in such simulations and tends to capture only a shorter segment compared to the orbital evolution duration of the binary system.

Multiple numerical hydrodynamic simulations have suggested that CBDs form after the CE evolution stage \citep{DM11, I11, KS11, S98}.
However, calculations consistent with the post-CE CBD evolution from the beginning of CE mass ejection are quite difficult.
Numerical simulations focusing on the interaction between the CBD and the binary system have also been performed.
Results suggest that CBD tides contract orbital separation and slightly increase the orbital eccentricity \citep{A91, SWH23}.
On the contrary, in some conditions, it has also been reported that angular momentum is transferred from the CBD to the binary star and the binary orbit may become larger \citep{MSS19}.
The modeling results of such interactions between circumbinary matter and the binary are highly dependent on the resolution of the simulations \citep{GP25, Th17}, and a correct estimation of the torque involves the accurate treatment of sink particles \citep{DR21, D24}.
In addition, numerical simulations have not yet been performed for such a sufficiently long duration for the disk to fully dissipate.

In binary evolution research, population synthesis studies have been conducted using the results of post-CE orbits calculated in the energy formalism ($\alpha \lambda$ formalism) \citep{B02, HTP02}.
In this form, the binary orbit after CE mass ejection is given by assuming that a fraction of the orbital kinetic energy $\alpha_{\rm{CE}}$ is used to overcome the binding energy of the envelope \citep{W84}.
Estimates based on such population synthesis play a major role in the estimation of the coalescence rate of gravitational wave sources.
However, binary evolution studies using this energy formalism have not been able to incorporate the effects of the post-CE CBD.

Herein, we analyzed a tight binary system after the CE evolution stage.
In such a post-CE binary system, a part of the outer layer ejected during CE evolution remains as remnant matter \citep{S98}.
If the remnant matter forms a CBD around the post-CE binary and interacts with it, the CBD can significantly impact the orbital evolution of the binary system.
In this study, we modeled the interaction between the post-CE binary system and its CBD and showed that the CBD contracts the binary orbit.
In particular, the orbit can shrink by up to $\sim 17$ \% after CE ejection.
Our simplified estimations incur no computational costs.
Further, incorporating them into population synthesis enables the study of evolution statistics of binaries with CBDs.

\section{Model}\label{sec:2}

In this study, the standard disk model was used to simulate the CBD \citep{SS73, ST83}.
The timescale for the contraction of the binary orbit $r$ due to the interaction of a binary with the CBD can be written as 
\begin{equation}
\frac{dr}{dt} = - \frac{r}{\tau_{\nu}} ,
\label{eq:evo}
\end{equation}
\noindent where  $\tau_{\nu}$ denotes the viscous timescale at the inner edge of the disk \citep{H09}
and the inner-edge radius of the disk is assumed to be identical to the orbital radius.

The standard accretion disk can be divided into three regions \citep{ST83}, and the viscous timescale depends on the region of the disk.
Following the formulation reported by Haiman, Kocsis and Menou (\yearcite{H09}) (see also \cite{H22}), we used the following viscous timescales for each region.  \\
(1) \textit{Inner region}: \\
In this region, radiation pressure dominates gas pressure and electron-scattering opacity is predominant 
\begin{equation}
 \tau_{\nu} = A \alpha_{\rm{vis}}^{-5/8} \dot{m}^{-13/8} M^{5/8} q_{\rm{s}}^{3/8} r^{35/16} .
 \label{eq:taunu1}
\end{equation}
(2) \textit{Middle region}: \\
In this region, gas pressure dominates radiation pressure and electron-scattering opacity is predominant
\begin{equation}
\tau_{\nu} = B \alpha_{\rm{vis}}^{-1/2} \dot{m}^{-5/8} M^{3/4} q_{\rm{s}}^{3/8} r^{7/8} .
\label{eq:taunu2}
\end{equation}
(3) \textit{Outer region}: \\
Here, gas pressure and free–free opacity are predominant,
\begin{equation}
\tau_{\nu} = C \alpha_{\rm{vis}}^{-8/17} \dot{m}^{-10/17} M^{12/17} q_{\rm{s}}^{7/17} r^{25/34} .
\label{eq:taunu3}
\end{equation}
Here, $A$, $B$, and $C$ are numerical coefficients \citep{H09}.
Further, $q_{\rm{s}} = 4q / \left( 1 + q^2 \right)$ is the normalized symmetric mass ratio, where $q = M_{2} / M_{1}$.
$M = M_{1} + M_{2}$ is the total mass of the binary ($M_1$ and $M_2$ denote the mass of the primary and companion stars, respectively).
$\dot{m}$ denotes the mass transfer rate in the CBD.
$\alpha_{\rm{vis}}$ is the viscous parameter of the standard disk model \citep{SS73}.

$\tau_{\nu}$ depends on the disk radius $r$ (proportional to as $r^{\beta}$).
If all dependencies except $r$ are compressed into $\tilde{\tau}_{\nu}$, the timescale $\tau_{\nu}$ can be written as 
$\tau_{\nu} = \tilde{\tau}_{\nu} r^{\beta}$.
From Equation~(\ref{eq:evo}), we obtain
\begin{equation}
\frac{dr}{dt} = - \frac{r^{1-\beta}}{\tilde{\tau}_{\nu}}.
\label{eq:drdt}
\end{equation}
Because the disk dissipates within a span of about the viscous time, integrating Equation~(\ref{eq:drdt}) over the initial viscous time $\tau_{\nu , 0} = \tilde{\tau}_{\nu} r_{\rm{ini}}^{\beta}$ results in 
\begin{equation}
\int_{r_{\rm{ini}}}^{r_{\rm{fin}}} r^{\beta-1}  dr = - \int_{0}^{\tau_{\nu, 0}} \frac{dt}{\tilde{\tau}_{\nu}} .
\label{eq:integ}
\end{equation}
We further obtain
\begin{equation}
\frac{1}{\beta} \left( r_{\rm{fin}}^{\beta} - r_{\rm{ini}}^{\beta} \right) = - \frac{\tau_{\nu,0}}{\tilde{\tau}_{\nu}} = - r_{\rm{ini}}^{\beta} ,
\end{equation} 
\noindent
where $r_{\rm{ini}}$ and $r_{\rm{fin}}$ denote the orbital radius just after CE ejection and at the end of CBD orbital evolution, respectively. 
Finally, we obtain
\begin{equation}
r_{\rm{fin}} = \left( 1 - \beta \right)^{1/ \beta} r_{\rm{ini}} .
\label{eq:rfin}
\end{equation}
\noindent
Here, we consider the inner-edge radius of the disk and the orbital separation of the binary to be equivalent \citep{H09}.
This indicates that the orbit of the binary shrinks by a factor $\left( 1 - \beta \right)^{1 / \beta} $ compared to the orbit just after CE evolution due to the effect of the CBD.
This factor is determined only by the $r$-dependent exponent of the viscous time and hence depends only on the region of the disk considered (inner, middle, or outer).

In most cases, the binary orbit lies in the outer region of the CBD, which gives $\beta = 25/34$.
Consequently, the post-CE CBD leads to the contraction of the orbital separation by a factor of $\left( 1 - \beta \right)^{1 / \beta} \sim 0.17$.
Here, all physical quantities of the disk other than the radius are merged into $\tilde{\tau}_{\nu}$ and do not explicitly appear in the results.
Notably, it is assumed that there is no external mass supply and the disk dissipates within the viscous timescale.
For such isolated CBDs, the contraction rate of the orbital separation is shown to be 
\begin{equation}
\frac{r_{\rm{fin}}}{r_{\rm{ini}}} = \left( 1 - \beta \right)^{1 / \beta} \sim 0.17.
\end{equation}
Actually, the viscous time $\tau_{\nu}$ may depend on $\dot{m}$, $q_{\rm{s}}$, and $\alpha_{\rm{vis}}$.
If the viscous time is short, the orbital contraction of the binary system is rapid.
However, this effect is limited because the disk dissipates quickly.
This correlation between orbital contraction time and disk disappearance time could potentially cancel out changes in viscous time.
Meanwhile, if the viscous time is long, the orbital contraction of the binary system proceeds slowly and the disk lives longer, which may lead to the same effect.

Unger, Grichener, and Soker (\yearcite{UGS24}) evaluated the impact of CBDs by incorporating a model that parameterized the transfer rate of orbital angular momentum into the envelope in their population synthesis computations and reported that the CBD effect was significant in post-CE binary systems containing black holes.
Their results showed that for double black hole and black hole-neutron star binaries, the peak orbital separation is $\sim 20\%$ lower when the CBD is present compared to when it is absent (see Figures 3 and 4 in \cite{UGS24}).
This result is very similar to the value obtained in our study ( 17\% ).
\citet{Va24} studied binary evolution in the presence of a CBD by interpolating existing numerical simulation results.
According to their model, although the disk dissipates faster than binary orbital evolution when the CBD mass is low, the orbital separation shrinks to a fraction of its post-CE value due to the presence of a CBD when the disk mass is sufficiently high.

\begin{figure}
 \begin{center}
  \includegraphics[width=80mm]{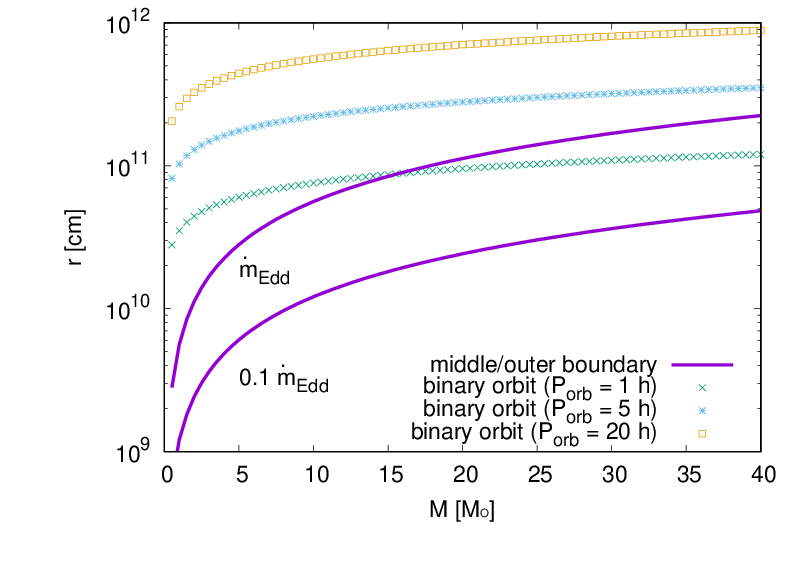}
 \end{center}
 \caption{ 
The relationship between the binary orbital radii and the disk boundary radii.
For orbital periods of 1, 5, and 20 h, the sequence of points represents the orbital separation as a function of the total binary mass.
Meanwhile, the boundary between the middle and outer regions of the standard disk is shown as a curve for the corresponding binary masses.
The lower curve corresponds to a mass transfer rate within the disk of $\dot{m} = 0.1 \dot{m}_{\rm{Edd}}$.
Meanwhile, the upper curve corresponds to $\dot{m} = \dot{m}_{\rm{Edd}}$.
The figure shows that the binary orbit mostly lies in the outer region.
However, the binary orbit may enter the middle region only when the orbital period is significantly short, the disk mass transfer rate is large, and the binary mass is high.

 }\label{fig:region}
\end{figure}

If a binary system contains a compact companion and has a tight orbit, the inner-edge of the disk may lie in the middle region.
In this case, $\beta = 7/8$ and the orbital separation shrinks by a factor of 0.093.
However, the disk can penetrate into the middle region in only extremely tight systems or when the mass transfer rate is very high in the disk.
Figure~\ref{fig:region} shows the boundary of the middle and outer regions with respect to the total mass of the binary system.
In this figure, the orbital separations corresponding to binary orbital periods of 1, 5 and 20 h are plotted as a sequence of points.
The boundary between the middle and outer regions is indicated by two curves.
The upper curve shows the case where $\dot{m}$ corresponds to the Eddington limit $\dot{m}_{\rm{Edd}}$, and the lower curve shows the case where $\dot{m} = 0.1 \dot{m}_{\rm{Edd}}$.
The upper sections of these curves correspond to the outer region of the disk.
The orbits can contract to the middle region only in binary systems with large total masses, extremely short orbital periods, and high mass transfer rates.
Such systems with narrow orbits are likely to coalesce through CBD evolution.
Hence, in most cases, only the outer region must be considered in analysis.

\section{Discussion}\label{sec:3}

In the current model, all disk and binary parameters are integrated into $\tilde{\tau}_{\nu}$, and this value is treated as a constant.
As expressed in Equations~(\ref{eq:taunu1}) - (\ref{eq:taunu3}) (see also Equation (26) in \cite{H09}), the viscous timescale depends on the total mass and the mass ratio of the central binary system.
During CBD evolution, the mass variation of binary components may be negligible.
This is because the evolution time limited by the viscous time is 10 years to a few thousand years at most.
Hence, even if the disk matter accretes onto the central object at the Eddington rate, the mass variation should be small.  
Due to the same reason, the mass ratio of the binary can be constant.

Meanwhile, $\tilde{\tau}_{\nu}$ includes the viscosity parameter $\alpha_{\rm{vis}}$ and the mass transfer rate $\dot{m}$.
In the outer region of the disk, $\tilde{\tau}_{\nu}$ is dependent on $\alpha_{\rm{vis}}^{-8/17}$.
Although $\alpha_{\rm{vis}}$ depends on complex microphysics in the disk, it is expected to take a value in the range of $0.1 - 1$ \citep{ST83, KPL07}.
The value $0.1$ appears to be commonly used in numerical studies concerning CBDs \citep{SWH23,Va24}. 
The 2D simulation reported by Thun, Kley, and Picogna (\yearcite{Th17}) is one of the cases using the value 0.01.
Considering various factors, the variation in $\tilde{\tau}_{\nu}$ resulting from changes in $\alpha_{\rm{vis}}$ is expected to be limited to a few times at most.
The mass transfer rate $\dot{m}$ may vary to some extent, although there is limited information available about the $\dot{m}$ value for the CBD after CE evolution.
It would not be surprising for $\dot{m}$ to exhibit fluctuations of $1 - 2$ orders of magnitude over short timescales, as observed for various accretion disk systems.
Numerical simulations of CBD accretion onto a binary black hole system showed that the mass transfer rate (matter supplied from the disk to the binary) exhibited oscillations of $1 - 2$ orders of magnitude coupled to the binary period, but did not appear to exhibit extreme changes \citep{DHM13, F14}.
Without external mass supply to the disk, the mass transport rate would naturally decrease in the long term.
However, in the outer region $\dot{m}$ is dependent on $\alpha_{\rm{vis}}^{-10/17}$, which does not generate large variation.
Therefore, we can assume that neither $\dot{m}$ nor $\alpha_{\rm{vis}}$ would be significantly affected when merged into $\tilde{\tau}_{\nu}$.
Moreover, as discussed earlier, $\tilde{\tau}_{\nu}$ controls the duration of integration expressed in Eq.~(\ref{eq:integ}), indicating that when $\tilde{\tau}_{\nu}$ becomes large, the integration time extends while orbital evolution slows down.
Hence, fluctuations in $\tilde{\tau}_{\nu}$ can somehow be negated. 
Also \citet{T25} suggested that CBDs reaching equilibrium may lead to stable angular momentum extraction persisting throughout the viscous time.

In most studies, population synthesis treated the orbital evolution of the binary before and after CE evolution using the energy formalism described below \citep{W84}.
Assuming that the initial orbital separation at the start of CE evolution is $a_0$ and the final separation after CE evolution is $a$, the change in the orbital kinetic energy before and after CE evolution is 
\begin{equation}
\Delta E_{\rm{orb}} = \frac{G M_{2}}{2} \left( \frac{M_{\rm{c}}}{a} - \frac{M_{1}}{a_{0}} \right) .
\label{eq:eorb}
\end{equation}
Here, $M_{\rm{c}}$ denotes the core mass of the primary star.
The binding energy of the giant star envelope stripped away through CE evolution can be expressed as 
\begin{equation}
E_{\rm{bind}} = \frac{G M_{1} \left( M_{1} - M_{\rm{c}} \right) }{\lambda R_{1}} .
\label{eq:ebind}
\end{equation}
$R_{1}$ denotes the radius of the primary star.
Here, if only a fraction $\alpha_{\rm{CE}}$ of the orbital energy is used to strip off the stellar envelope, then
\begin{equation}
\alpha_{\rm{CE}} \Delta E_{\rm{orb}} = E_{\rm{bind}}.  \nonumber
\end{equation}
That is,
\begin{equation}
\frac{ G M_{2}}{2} \left( \frac{M_{\rm{c}}}{a} - \frac{M_{1}}{a_{0}} \right) \alpha_{\rm{CE}} \lambda 
= \frac{G M_{1} \left( M_{1} - M_{\rm{c}} \right) }{R_{1}} .
\label{eq:al}
\end{equation} 
Usually, this equation is solved to obtain post-CE orbital separation $a$.
Here, $\lambda$ denotes the central concentration of the giant star, which always appears with $\alpha_{\rm{CE}}$.
Therefore, its effect is absorbed into $\alpha_{\rm{CE}}$, i.e., we take $\lambda = 1$.
The $a$ value obtained in this way is the orbital separation after CE evolution.
However, if the CBD remains after the CE evolution phase, the orbit shrinks further and the binary orbit eventually contracts to $a^{\prime} = \left( 1 - \beta \right)^{1/ \beta} a$.
This relation is obtained by assuming that the inner-edge radius of the disk and the orbital separation of the binary are equivalent \citep{H09}, and rewriting the disk inner radius $r$ to the orbital separation $a$ in Equation~(\ref{eq:rfin}).
This relation can be used to handle CE and CBD evolution simultaneously and determine the effective energy fraction consumed for stripping the envelope ( $\alpha_{\rm{CE+CBD}}$ ).
By dividing this effective energy fraction by $\alpha_{\rm{CE}}$, Equation~(\ref{eq:al}) is solved for determining the fraction of the orbital energy
\begin{eqnarray}
\gamma &=& \frac{\alpha_{\rm{CE+CBD}}}{\alpha_{\rm{CE}}} \nonumber \\
&=& \frac{G M_{1} \left( M_{1} - M_{\rm{c}} \right) }{\alpha_{\rm{CE}} R_{1}} 
\left[ \frac{ G M_{2}}{2} \left( \frac{M_{\rm{c}}}{a^{\prime}} - \frac{M_{1}}{a_{0}} \right) \right]^{-1} ,
\label{eq:gammanew}
\end{eqnarray} \noindent
or by using explicit binary parameters and $\beta$, 
\begin{eqnarray}
\gamma 
&=&  \frac{      
\left[ \frac{G M_{\rm{c}} M_{2}}{2 a'} - \frac{G M_{1} M_{2}}{2 a_{0}} \right]^{-1}   
}{
\left[ \frac{G M_{\rm{c}} M_{2}}{2 a} - \frac{G M_{1} M_{2}}{2 a_{0}} \right]^{-1}   
} \nonumber \\
&=& \frac{1  - M_{1} a / \left( M_{\rm{c}} a_{0}  \right) }{ \left( 1 - \beta  \right)^{-1/ \beta} - M_{1} a / \left( M_{\rm{c}} a_{0}  \right) }  .
\label{eq:gamma}
\end{eqnarray}
\noindent If $M_1$ and $M_{\rm{c}}$ are determined, $a$ can be obtained from Equation~(\ref{eq:al}).
$\alpha_{\rm{CE+CBD}}$ is a constant determined by $\beta$.
Therefore, if $\alpha_{\rm{CE+CBD}}$ is obtained in advance and used instead of $\alpha_{\rm{CE}}$, the effect of the CBD can be incorporated without changing the conventional population synthesis code.
Assuming that regular CE evolution results in $a \ll a_0$, the following equation is obtained:
\begin{equation}
\alpha_{\rm{CE+CBD}} \simeq \left( 1 - \beta \right)^{1/\beta} \alpha_{\rm{CE}} .
\label{eq:approxalpha}
\end{equation}

\begin{figure}
 \begin{center}
  \includegraphics[width=80mm]{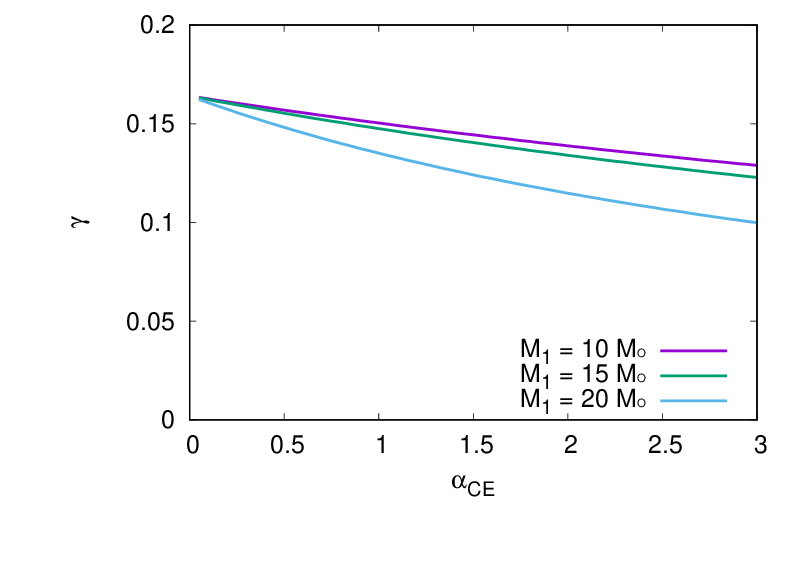}
 \end{center}
 \caption{ 
$\gamma$ as a function of $\alpha_{\rm{CE}}$ computed using Eq.~(\ref{eq:gammanew}).
Results for three mass values ($10 - 20 \rm{M}_{\odot}$) are shown.
When $\alpha_{\rm{CE}}$ is small, that is, when the orbit has become sufficiently small through CE evolution, the approximation in Eq.~(\ref{eq:approxalpha}) holds well.
In other words, over a reasonably wide range of parameters, it is reasonable to multiply the value of $\alpha_{\rm{CE}}$ by 0.17 to account for the effect of the CBD.
 }\label{fig:gamma}
\end{figure}

Unfortunately, since $ \left( M_{1} a \right) / \left( M_{\rm{c}} a_{0} \right) $ is not always negligible, we need to consider more plausible values of orbital radii and masses.
Figure\ref{fig:gamma} shows the variation of $\gamma$ with respect to $\alpha_{\rm{CE}}$.
To estimate $\gamma$, the primary star is assumed to be a giant star with an initial mass of 10, 15, and 20 $\rm{M}_{\odot}$ and the companion star is assumed to be a compact object with a mass of $1.4 \rm{M}_{\odot}$.
The approximated evolution of individual massive stars is computed to determine their mass, radius, and core mass at the end of their evolution sequence \citep{H00, Karino20}.
Using these parameters, CE evolution with an efficiency of $\alpha_{\rm{CE}}$ is modeled, followed by CBD evolution.
As shown in Figure \ref{fig:gamma}, for primary stars with masses of $10$ and $15 \rm{M}_{\odot}$, $\alpha_{\rm{CE+CBD}} \simeq \left( 1 - \beta \right)^{1/\beta} \alpha_{\rm{CE}}$ is a very good approximation.
However, the approximation is somehow worse for $20 \rm{M}_{\odot}$.
This is because stellar winds become strong and most of the outer layers are ejected before CE evolution for such massive stars.
Further, the orbital separation after CE evolution increased (Equation.~(\ref{eq:gamma})).

The current model may be oversimplified in some respects and requires examination for its applicability.
Nevertheless, the result that the orbital radius shrinks further by up to $\sim 17$ \% due to the presence of the CBD after CE evolution is highly significant.
This result is applicable -- for example -- to double neutron star binaries; hence, it will affect their merger rate.
Previous population synthesis studies suggested that a double neutron star binary merging within the cosmic age would occur only when a second neutron star is born in a fairly tight binary \citep{B02}.
However, if the CBD causes orbital contraction, tight systems forming during CE evolution are likely to merge just after the CE evolution phase.
In a binary system consisting of a massive stellar core and a neutron star, the binary will merge before the core undergoes a supernova explosion and becomes a neutron star.
Consequently, in tight binaries, the formation of a double neutron star may be impossible.
Systems with wide orbits may show orbital contraction during the CBD evolution phase that follows the CE evolution phase and act as the progenitors of merging double neutron stars.
Consequently, further studies are required to accurately evaluate the effects of CBDs on binary evolution and realize population synthesis that considers the impacts of CBDs.
In such population synthesis studies, the approximated formulation given in Equations~(\ref{eq:gamma}) and (\ref{eq:approxalpha}) will be useful.

For the CBD to extract angular momentum, a sufficiently massive disk must remain following CE evolution.
In the widely used $\alpha \lambda$ formalism, 
the binding energy required to strip away the outer layers of a giant star is given by Equation~(\ref{eq:ebind}).
However, this binding energy is the energy required to strip the outer envelope away from the giant star whose mass is $M_{1}$.
The energy equivalent to this binding energy does not imply that it is sufficient to strip the outer layers from the entire binary system.
Therefore, if the orbit of the binary system was narrowed to the point predicted by the $\alpha \lambda$ model via CE evolution, a significant portion of the stripped mass of the outer layer ( $M_{1} - M_{\rm{c}}$ ) would remain bound to the binary system.
Hence, it is natural to assume that a disk with enough mass to affect the binary orbit will be created.

Although the above discussion is primitive, more precise numerical calculations indeed suggest the existence of a CBD.
According to 1D simulations by \citet{F19}, after the CE evolution of a massive star and a neutron star, a thin outer envelope remains around the massive star core, suggesting stable Roche-lobe overflow (RLOF) onto the neutron star (see also \cite{N25}). 
In this case, RLOF and CBD evolution could occur simultaneously. 
Since CBD evolution proceeds on a shorter timescale than stable RLOF, the binary orbit would rapidly contract. 
Consequently, the neutron star might plunge into the remaining envelope, initiating further CE evolution. 
In such a scenario, due to the initial orbit is small, the system would ultimately merge.
Large-scale 3D numerical calculations indicate that CE evolution does not progress in a spherical manner.
Rather, the outer layers of the giant star show a tendency to expand in a disk-like manner \citep{V24}.
Additionally, incorporating the effect of radiative diffusion significantly impedes the ejection of the outer layers. 
It has been demonstrated that even after the plunge-in phase terminates, a considerable portion of the outer envelope remains around the binary system \citep{L25}.
If the CBD formed in this manner extracts angular momentum from the binary system, the effective value of $\alpha_{\rm{CE}}$ will decrease. Consequently, after CE evolution followed by CBD evolution, a tight binary system will remain.

Numerical simulations of CBD--binary interactions have revealed the impact of the orbital eccentricity of the binary system on binary evolution \citep{SWH23}.
When the eccentricity is low, interactions of the system with the disk may cause the orbital separation to increase as the eccentricity increases.
However, even in such cases, once the eccentricity is saturated, the orbital separation transitions into orbital contraction.
For simplicity in this study, the orbital eccentricity was set to zero; however, this aspect may need to be considered carefully in future studies. 
When the central binary system is close to a circular orbit, accretion from the disk concentrates onto the lighter star \citep{AL96}. 
The concentration of mass and angular momentum on the lighter companion may cause an increase in orbital eccentricity. 
This increase in eccentricity would temporarily push the inner edge of the disk outward. 
Subsequently, a balance develops between tidal removal of angular momentum from the binary and accretion from the CBD. 
If the disk is sufficiently massive, its increased density \citep{LP74} will eventually push the binary orbit inward. 
However, more detailed studies are needed to reveal the long-term evolution of the binary system driven by the CBD.

In the simple energy formalism ($\alpha \lambda$ formalism) \citep{W84}, $\alpha_{\rm{CE}}$ never exceeds unity under the assumption that the energy is supplied only from the orbital energy. 
However, if other energy sources are available, $\alpha_{\rm{CE}}$ can effectively exceed unity.
For example, recombination energy in the outer layers of giant stars can substantially increase $\alpha_{\rm{CE}}$ \citep{I13}.
$\alpha_{\rm{CE}}$ can also increase due to the conversion of magnetic energy into mechanical work.
In the three-dimensional magnetohydrodynamic simulation of stellar mergers performed by \citet{S19}, rapid amplification of the initial magnetic field was observed during the merger.
Even in the case of CE evolution, it has been noted that a weak initial magnetic field can be amplified by several orders of magnitude due to magneto-rotational instability \citep{O16}.
The amplification of the magnetic field may lead to outflow from the outer layer, such as a high-velocity jet stream, in the polar direction \citep{V24}.
It may further contract the binary orbit through angular momentum transfer via magnetic fields \citep{WMH25}.

From the observational perspective, the fraction of double neutron star binaries observed currently does not agree with the predictions of population synthesis studies considering small $\alpha_{\rm{CE}}$ values.
To reproduce the observational results, it is suggested that the merger rate of Hertzsprung-gap donors during CE evolution must be increased and the value of $\alpha_{\rm{CE}}$ must be larger \citep{C25}.
For close binary systems involving main-sequence stars and white dwarfs, it has been suggested that a small value of $\alpha_{\rm{CE}}$~($\lesssim 0.3$) is favorable for reproducing the observed orbital period distribution \citep{S25}.
The value of $\alpha_{\rm{CE}}$ may vary depending on the mass and evolutionary stage of the companion \citep{DM11}.
However, the results obtained in this study suggest -- at the least -- that CBD evolution following CE ejection is universal.
To accurately quantify the residual envelope mass and assess its impact on the subsequent evolution of the binary system, further investigations integrating numerical simulations and observational data are essential.

\section{Conclusion}

CE evolution plays a significant role in binary evolution and is considered particularly important in generating compact binaries with short orbital periods.
In this study, we evaluated the impact of a CBD, which likely originates from CE ejection, on binary evolution following CE evolution.
The results indicated that the presence of a CBD can contract the orbits of binary systems after CE evolution by up to 17 \%.
This effect is considered highly universal and independent of the CBD's structure.
Although our analysis approach might be oversimplified, it incurs minimal computational costs.
Compared to numerical simulations, simple analytic methodologies possess distinct advantages. 
Advancing research using both methodologies will provide important insights into the evolution of post-CE systems and constitute an essential step in discussing compact binary formation and merger rates.
Considering the effect of a CBD on binary evolution, it may be necessary to re-examine the evolutionary processes of binary systems leading to the formation of double black hole and double neutron star binaries.





%


\end{document}